\documentclass[conference]{IEEEtran}
\usepackage[T1]{fontenc}
\usepackage[utf8]{inputenc}
\usepackage[english]{babel}
\usepackage{algorithm}
\usepackage[noend]{algpseudocode}
\usepackage{cite}
\usepackage{amsmath, mathtools}
\usepackage{amsfonts, amsthm, bm}
\usepackage{amssymb}
\usepackage{graphicx}
\usepackage[acronym,shortcuts]{glossaries}
\usepackage{balance}

\algnewcommand{\IfThen}[2]{
  \State \algorithmicif\ #1\ \algorithmicthen\ #2}

\usepackage{tikz}
\usepackage{pgfplots}
\pgfplotsset{compat=newest}
\pgfplotsset{plot coordinates/math parser=false}
\newlength\fheight
\newlength\fwidth
\usetikzlibrary{plotmarks,patterns,decorations.pathreplacing,backgrounds,calc,arrows,arrows.meta,spy,matrix}
\usepgfplotslibrary{patchplots,groupplots}
\usepackage{tikzscale}

\usetikzlibrary {patterns.meta}

\DeclareMathOperator*{\argmin}{\arg\!\min}

\newacronym{3gpp}{3GPP}{3rd generation partnership project}
\newacronym{5g}{5G}{fifth-generation}
\newacronym{b5g}{B5G}{beyond-fifth-generation}
\newacronym{6g}{6G}{sixth-generation}
\newacronym{aiot}{A-IoT}{Ambient Internet of Things}
\newacronym{als}{ALS}{alternating least squares}
\newacronym{awgn}{AWGN}{additive white Gaussian noise}
\newacronym{bpsk}{BPSK}{binary phase-shift keying}
\newacronym{bs}{BS}{base station}
\newacronym{crdsa}{CRDSA}{contention resolution diversity slotted-ALOHA}
\newacronym{cpd}{CPD}{canonical polyadic decomposition}
\newacronym{csi}{CSI}{channel state information}
\newacronym{fec}{FEC}{forward error correction}
\newacronym{fsa}{FSA}{framed slotted-ALOHA}
\newacronym{iot}{IoT}{Internet-of-things}
\newacronym{irsa}{IRSA}{irregular repetition slotted-ALOHA}
\newacronym{lmmse}{LMMSE}{linear minimum mean square error}
\newacronym{mac}{MAC}{medium access control}
\newacronym{mer}{MER}{message error rate}
\newacronym{mra}{MRA}{massive random access}
\newacronym{mtc}{MTC}{machine-type communications}
\newacronym{nr}{NR}{new radio}
\newacronym{ook}{OOK}{on-off keying}
\newacronym{noma}{NOMA}{non-orthogonal multiple access}
\newacronym{ra}{RA}{random access}
\newacronym{saloha}{SALOHA}{slotted ALOHA}
\newacronym{sic}{SIC}{successive interference cancellation}
\newacronym{snr}{SNR}{signal-to-noise ratio}
\newacronym{ura}{URA}{unsourced random access}
\newacronym{tbra}{TBRA}{tensor-based random access}

\title{Tensor-based Random Access\\for Ambient IoT Contention Resolution}

\author{Alberto Rech, Alexis Decurninge, Ala Baccar, and Sofiane Kharbech\\
\small
Advanced Wireless Technologies Lab, Fourier Research Center, Huawei Technologies, Paris, France.\\
\texttt{name.surname@huawei.com}
}
\begin{document}
\maketitle

\begin{abstract}
\Ac{aiot} deployments impose stringent hardware constraints, including low-order modulations and simple transceiver architectures. We propose a tensor-based grant-free random access scheme for the initial access message (Msg1) tailored to these limitations. Each device transmits replicas of its short payload structured as a rank-1 tensor over multiple access occasions. The \ac{aiot} receiver performs joint activity detection, channel estimation, and decoding via tensor decomposition and successive interference cancellation. Results show significant throughput gains over conventional schemes while preserving low-complexity transmitter designs.
\end{abstract}

\begin{picture}(0,0)(0,-280)
\put(0,0){
\put(0,0){\qquad \qquad \quad This paper has been submitted to IEEE for publication. Copyright may change without notice.}}
\end{picture}


\glsresetall

\section{Introduction}

\label{sec:introduction}

\Ac{aiot}~\cite{Butt2024Ambient, Zheng2024Ambient} has recently emerged as a promising paradigm for enabling massive connectivity among ultra-low-power devices characterized by limited encoding and modulation capabilities. In the initial access phase of \ac{aiot} inventory procedures, a reader broadcasts a paging message, triggering uplink transmissions (Msg1) from all recipient devices. Upon reception of the paging message, a potentially massive number of devices transmit short identification sequences in an uncoordinated manner within a constrained time–frequency resource budget. This process is contention-based: devices transmit a random identifier in Msg1, followed by a dedicated contention-resolution phase to manage collisions.

Due to stringent hardware and energy constraints, only very low-complexity transmission schemes are feasible in practical \ac{aiot} deployments~\cite{Alnahari2025Ambient,Singh2025Physical}. In particular, devices are typically restricted to binary modulations such as \ac{bpsk} or \ac{ook}, and simple medium access schemes. As a result, \ac{saloha}, wherein each device selects a single access occasion to transmit its message, is typically adopted for Msg1 transmissions. While attractive for its simplicity, this approach suffers from severe collisions at moderate-to-high system load, leading to frequent message misdetections and poor scalability.

At the \ac{mac} layer, coded repetition schemes such as \ac{crdsa}~\cite{Casini2007Contention} and \ac{irsa}~\cite{Liva2011Graph}
use packet replication and \ac{sic} to mitigate collisions, improving throughput over \ac{saloha}. However, with these approaches, the repetitions do not follow any structured pattern, and each replica is treated independently at the receiver. 

Within the \ac{ura} framework~\cite{Polyanskiy2017A}, tensor-based modulation schemes~\cite{Decurninge2021Tensor, Rech2023Unsourced} instead exploit this structure to jointly perform activity detection and decoding, yet typically require large modulation orders, sophisticated \ac{fec}, and multi-antenna receivers, and do not leverage the discreteness of the constellations. Initial efforts in the latter direction include dictionary-based tensor decomposition~\cite{Cohen2018Dictionary} and a hybrid \ac{ura} design~\cite{Baccar2025A} by some of the authors of this letter; the latter, however, still relies on signal structures and transceiver complexities beyond the reach of low-cost \ac{aiot} devices.

Focusing on \ac{aiot} frameworks, two recent approaches have been proposed in ~\cite{Song2026Dynamic} and~\cite{Qiao2025A}. The former provides a  \ac{mac}-layer analysis but neglects modulation, coding constraints, and propagation effects. The latter exploits the superposition of \ac{ook} symbols to construct collision constellation diagrams and resolve second-order collisions. Although suitable for general random-access scenarios, this approach is not well adapted to payload-limited Msg1 transmissions, as it relies on preambles and training sequences for channel estimation.

In this letter, we introduce a grant-free \ac{tbra} scheme specifically designed for \ac{aiot} Msg1 transmissions. In contrast to classical repetition-based random access, the proposed approach introduces and exploits a multi-dimensional structure induced by replica transmissions: each device maps its low-order-modulated payload onto multiple access occasions, yielding a structured rank-1 tensor across time, frequency, and replica dimensions. At the \ac{aiot} reader, this structure is explicitly leveraged through a dictionary-based tensor decomposition algorithm combined with slot-level \ac{sic}, enabling joint activity detection, channel estimation, and decoding under non-coherent conditions.
Unlike existing tensor-based \ac{ura} schemes, the proposed design operates with binary modulation, single-antenna reception, and no reliance on \ac{fec}, making it fully compatible with practical \ac{aiot} constraints. At the same time, it differs from conventional repetition schemes by introducing and exploiting the algebraic coupling among replicas rather than treating them independently.

The rest of the letter is organized as follows. In Section~\ref{sec:system_model}, we introduce the system model. In Sections~\ref{sec:allocation} and ~\ref{sec:separation}, we present the proposed \ac{tbra} scheme transceiver design.
A performance comparison with state-of-the-art schemes and the conclusions are presented in  Sections~\ref{sec:numerical_results} and~\ref{sec:conclusions}, respectively.

\emph{Notation.} 
Scalars, column vectors, matrices, and tensors are denoted by italic, boldface lowercase, and boldface uppercase italic letters, and boldface upright letters, respectively. Sets are denoted by calligraphic uppercase letters, with $|\mathcal{A}|$ denoting the cardinality of set $\mathcal{A}$. $\bm{A}^{\rm T}$, $\bm{A}^{*}$ and $\bm{A}^\dagger$ denote the transpose, conjugate, and conjugate transpose of $\bm{A}$, respectively. The operators $\otimes$ and $\odot$ denote the Kronecker and Khatri-Rao products, respectively. $\mathbb{E}[\cdot]$ the statistical expectation.

\section{System Model}
\label{sec:system_model}

We consider a grant-free \ac{aiot} random access scenario in which $K_{\rm a}$ single-antenna active users transmit a random ID (\textit{payload}) of $B$ bits to a single-antenna receiver (\textit{\ac{aiot} reader}) within a random access frame composed of $N$ access occasions. Each occasion consists of $T$ time-frequency resource elements.
\footnote{This setup is specifically aligned with the "Solution 1" architecture depicted in \cite[Sec.~6.3.4]{3gpp38769}. The \ac{aiot} device does not utilize a unique, pre-assigned hardware identity for random access. Instead, the device generates a $16$-bit \textit{random ID} and transmits it in the \ac{aiot} Msg1 on the identified access occasion(s), with no \ac{fec} coding. We remark that this architectural choice is fundamentally aligned with the \ac{ura} framework; the proposed solution can thus be extended to any other system wherein short payloads, limited coding capabilities, and low-order modulation constraints are imposed.}
Each user transmits their symbol vector $\bm{s}_k \in  \mathbb{C}^{T}$ over a subset of access occasions within the frame. Let $\mathcal{S}_k \subseteq \{1,\dots,N\}$ denote the set of access occasions selected by user $k$. We assume a quasi-static flat-fading channel over the $TN$ time-frequency resources, with coefficient $h_k \sim \mathcal{CN}(0,1)$. The received signal in occasion $n$ is,
\begin{equation}
\bm{y}_n  = \sum_{k : n \in \mathcal{S}_k} h_k \bm{s}_k + \bm{w}_n, \quad n = 1,\dots,N,
\end{equation}
where $\bm{w}_n$ is the \ac{awgn} vector with power $\sigma_w^2$.
Stacking the received occasion vectors into a matrix 
$\bm{Y} = [\bm{y}_1,\dots,\bm{y}_N] \in \mathbb{C}^{T \times N}$ 
yields
\begin{equation}
\bm{Y} = \sum_{k=1}^{K_{\rm a}} h_k \bm{s}_k \bm{a}_k^{\rm T} + \bm{W},
\end{equation}
where $\bm{a}_k \in \{0,1\}^N$ is the occasion-indicator vector of user $k$, with $[\bm{a}_k]_n = 1$ if $n \in \mathcal{S}_k$ and $0$ otherwise, and $\bm{W}$ collects the noise vectors $\bm{w}_n$.
This formulation reveals a sum of rank-one components, which constitute the structural basis exploited by the proposed tensor-based receiver.

\section{Tensor-Based Random Access}
\label{sec:allocation}

This section presents the proposed \ac{tbra} scheme by describing the structured transmitter design that induces a separable multi-dimensional access pattern at the receiver side.

\subsection{Tensor Structure}
Assuming the number of access occasions $N$ is factorized by some $N_\ell$, $\ell \in \{1,\dots,L\}$, i.e., $N = \prod_{\ell=1}^{L} N_\ell$, and assuming there exist some $\bm{c}_{k,\ell}$ of size $1 \times N_{\ell}$ for all $\ell$ and $k$ such that
\begin{equation}
    \bm{a}_k = \left(\bm{c}_{k,1} \otimes \cdots \otimes \bm{c}_{k,L}\right)^{\rm T},
\end{equation}
for all $k$, matrix $\bm{Y}$ can be reshaped to an $(L+1)$-mode tensor. Specifically, the noiseless tensor $\mathbf{Y} \in \mathbb{C}^{T \times N_1 \times \cdots \times N_L}$ can be written as 
\begin{equation}\label{eq:tensor}
\mathbf{Y} = \sum_{k=1}^{K_{\rm a}} h_k\bm{s}_k \otimes \bm{c}_{k,1} \cdots \otimes \bm{c}_{k,L}.
\end{equation}

Therefore, the objective of the random access scheme is the joint design of the resource allocation components $\bm{c}_{k,\ell}$, for all $k$, $\ell$, and of a user separation algorithm at the receiver side, which allows for decomposing the tensor into the single user (rank-1) components.

\subsection{Transmitter Design \& Occasion Selection}

\begin{figure}[t]
    \centering
    \includegraphics[width=0.99\linewidth]{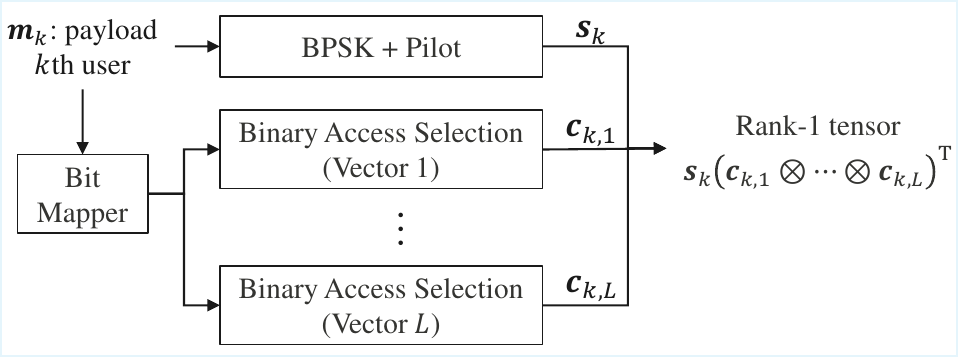}
    \caption{Transmitter tensor-based occasion selection.}
    \label{fig:transmitter}
\end{figure}

The transmitter architecture is designed to accommodate the stringent power and complexity constraints of \ac{aiot} devices. To avoid the significant overhead and hardware requirements associated with low-rate channel coding or long training sequences (e.g., as explored in \cite{Qiao2025A}), we exploit the inherent randomization of the transmitted payload as specified in \cite{3gpp38769}.

Specifically, each user $k$ maps its $B$-bit payload to a unique tuple of binary codewords $\{\bm{c}_{k,1}, \dots, \bm{c}_{k,L}\}$, where each $\bm{c}_{k,\ell}$ is drawn from a finite codebook $\mathcal{C}_\ell \subseteq \{0,1\}^{N_\ell}$ known at the receiver. 
Specifically, we impose that each access vector \eqref{eq:ck} is selected from a constant-weight codebook~\cite{Agostini2023Constant}, i.e., all the codewords in $\mathcal{C}_\ell$, share the same number of ones.
We define the deterministic mapping associating users' payloads (or subsets of their payloads) to codewords as
\begin{equation}\label{eq:ck}
\gamma(\bm{m}_k) = (\bm{c}_{k,1}, \dots, \bm{c}_{k,L}), \quad \bm{c}_{k, \ell} \in \mathcal{C}_\ell.
\end{equation}
This structured design provides a critical consistency check for the \ac{sic} process at the receiver side, as detailed in Section~\ref{sec:sic}. 
To facilitate both channel estimation and data recovery at the receiver, each active user appends a single pilot symbol to the encoded $B$-bit payload.
The $B$ bits in the payload are modulated using \ac{bpsk}, yielding a transmitted symbol vector $\bm{s}_k \in \{+1,-1\}^{T}$. 
The transmitter design scheme is reported in Fig.~\ref{fig:transmitter}. As the map $\gamma(\cdot)$ is known at the receiver, the integrity of a decoded user is verified at each iteration by checking if the recovered codewords belong to the predefined set of valid tuples, thereby distinguishing decoded signals successfully from residual interference.

\section{CPD-based User Separation}
\label{sec:separation}
The receiver exploits the separable structure induced at the transmitter and formulates multi-user detection as a constrained tensor factorization problem. Recalling that the received signal admits the $(L+1)$-mode tensor representation $\mathbf{Y} \in \mathbb{C}^{T \times N_1 \times \cdots \times N_L}$, the objective is to recover the rank-1 components corresponding to active users.
To this end, the receiver computes a constrained \ac{cpd} by solving
\begin{subequations}\label{eq:optproblem}
    \begin{IEEEeqnarray}{rCl}
    &\argmin_{\{\hat{\bm{x}}_k,\hat{\bm{c}}_{k,\ell}\}} & \left\| \mathbf{Y} - \sum_{k=1}^{\hat{K}} \hat{\bm{x}}_k \otimes \hat{\bm{c}}_{k,1} \otimes \cdots \otimes \hat{\bm{c}}_{k,L} \right\|_F^2,\\
    & \text{s.t.}\; & \hat{\bm{c}}_{k,\ell} \in \mathcal{C}_\ell, \quad \forall \ell, k\\
    & &\hat{\bm{x}}_k\in \mathbb{C}^T, \quad \forall k
    \end{IEEEeqnarray}
\end{subequations}
where $\hat{K}$ is an upper bound on the number of active users and $\hat{\bm{x}}_k \approx h_k \bm{s}_k$  captures the effective received signal.
Identifiability bounds for the rank-1 components were derived in~\cite{Decurninge2021Tensor} for the unconstrained continuous case; characterizing the identifiable region under the discrete codebook constraints is left for future work.  

\subsection{User Separation Through ALS}

Problem~\eqref{eq:optproblem} is addressed via \ac{als}, which cyclically updates one factor by optimizing \eqref{eq:optproblem} while keeping the others fixed, yielding a sequence of least-squares subproblems whose complexity scales linearly with the tensor dimensions~\cite{Kolda2009Tensor}. Collecting the factors into $\hat{\bm{X}}=[\hat{\bm{x}}_1,\dots,\hat{\bm{x}}_{\hat{K}}]$ and $\hat{\bm{C}}_\ell=[\hat{\bm{c}}_{1,\ell}^{\rm T},\dots,\hat{\bm{c}}_{\hat{K},\ell}^{\rm T}]$, and denoting by $\bm{Y}_{(m)}$ the mode-$m$ unfolding of tensor $\mathbf{Y}$, the update of the $m$-th factor reads
\begin{equation}\label{eq:als}
\bm{F}^{(m)} \leftarrow
\bm{Y}_{(m)}\,\bar{\bm{F}}^{(m)}
\big(\bar{\bm{F}}^{(m)\dagger}\bar{\bm{F}}^{(m)}+\lambda\bm{I}_{\hat{K}}\big)^{-1},
\end{equation}
where $\bm{F}^{(0)}=\hat{\bm{X}}$, $\bm{F}^{(\ell)}=\hat{\bm{C}}_\ell$, and $\bar{\bm{F}}^{(m)}=\bigodot_{j\neq m}\bm{F}^{(j)},$ is the Khatri-Rao product of all but the $m$-th tensor components. 
Note that, as the number of active users grows, the Khatri-Rao Gram matrix $\bar{\bm{F}}^{(m)\dagger}\bar{\bm{F}}^{(m)}$ becomes ill-conditioned, i.e., users' signatures overlap, and one or more of the Gram matrix's eigenvalues approach zero, making the matrix nearly non-invertible. Therefore, a fixed Tikhonov regularization term $\lambda$ is adopted to stabilize the update at high load.

After each update \eqref{eq:als}, the columns of the allocation-mode factors are rescaled to the codeword norm and constrained to be non-negative,
\begin{equation}\label{eq:norm}
\hat{\bm{c}}_{k,\ell} \leftarrow
\rho_\ell\,\frac{[\hat{\bm{c}}_{k,\ell}]_+}{\big\|[\hat{\bm{c}}_{k,\ell}]_+\big\|_2},
\end{equation}
where  $[\,\cdot\,]_+ \triangleq \max\!\big(\Re\{\cdot\},0\big)$ and $\rho_\ell=\|\bm{c}\|_2$ is constant over $\bm{c}\in\mathcal{C}_\ell$ since the codewords share the same Hamming weight. This rescaling fixes the per-iteration scaling freedom of the \ac{cpd} across modes, keeping the estimates commensurate with the binary codewords, while the non-negativity reflects the $\{0,1\}$ alphabet; together they render the nearest-codeword distances in~\eqref{eq:proj-c} meaningful.

The discrete structure of the access patterns is enforced via projection onto the codebooks. As in~\cite{Baccar2025A}, the continuous estimates of the allocation modes are periodically mapped to the nearest codeword,
\begin{equation}\label{eq:proj-c}
\Pi_{\mathcal{C}_\ell}(\hat{\bm{c}}_{k,\ell})
=\argmin_{\bm{c}\in\mathcal{C}_\ell}\big\|\bm{c}-\hat{\bm{c}}_{k,\ell}\big\|_2 .
\end{equation}
However, in contrast to~\cite{Baccar2025A}, where mapping is applied only to one of the tensor modes, the relatively small codebook sizes let us project across \emph{all} allocation dimensions, enforcing the discrete constraint jointly and improving the robustness of the recovered components.

Rather than snapping every mode simultaneously, the projections are staggered: after an initial warm-up of $I_{\rm W}$ sweeps that lets the factors settle in the continuous space, mode $\ell$ first projects at offset $\phi_\ell$ and repeats every $I_{\rm D}$ iterations. Each projection is followed by a short freeze of $I_{\rm F}$ iterations during which the snapped mode is held fixed while the remaining modes adapt to it. 
This gradual, mode-by-mode projection addresses the instability of enforcing all discrete constraints at once and lets confidently recovered codewords guide the estimation of the others. Differently, the signal mode $\hat{\bm{x}}_k\approx h_k\bm{s}_k$ is left as an unconstrained least-squares factor throughout the iterations, and its \ac{bpsk}-derived structure is instead exploited for channel estimation purposes only, as described in Section~\ref{sec:chest}. 
The overall procedure, which repeats the presented steps for $I_{\rm ALS}$ iterations, is summarized in Algorithm~\ref{alg:als}.

\begin{algorithm}[t]
\caption{Constrained \ac{cpd}-based User Separation}
\label{alg:als}
\begin{algorithmic}[1]
\Require $\mathbf{Y}$; $\hat{K}$; $\{\mathcal{C}_\ell\}_{\ell=1}^{L}$; $\lambda$;
         $I_{\rm ALS}$; $I_{\rm W}$; $\{\phi_\ell\}_{\ell=1}^{L}$, $I_{\rm D}$, $I_{\rm F}$
\Ensure factors $\hat{\bm{X}},\hat{\bm{C}}_1,\dots,\hat{\bm{C}}_L$
\State randomly initialize $\hat{\bm{X}}$, $\hat{\bm{C}}_\ell$
\For{$i = 1$ \textbf{to} $I_{\rm ALS}$}
  \For{$\ell = 0$ \textbf{to} $L$}
    \IfThen{mode $\ell$ is frozen}{\textbf{continue}}
    \State update $\bm{F}^{(\ell)}$ via \eqref{eq:als}
    \If{$\ell \ge 1$}
      \State normalize $\hat{\bm{C}}_\ell$ to codeword norm via \eqref{eq:norm}
      \If{$i > I_{\rm W}$ \textbf{\&} $(i\!-\!I_{\rm W}\!-\!\phi_\ell)\bmod I_{\rm D}=0$}
        \State $\hat{\bm{C}}_\ell \gets \Pi_{\mathcal{C}_\ell}(\hat{\bm{C}}_\ell)$ via \eqref{eq:proj-c}
        \State freeze mode $\ell$ for $I_{\rm F}$ iterations
      \EndIf
    \EndIf
  \EndFor
\EndFor
\State $\hat{\bm{C}}_\ell \gets \Pi_{\mathcal{C}_\ell}(\hat{\bm{C}}_\ell),\ \forall \ell$
\end{algorithmic}
\end{algorithm}

\subsection{Channel Estimation \& Message Decoding}
\label{sec:chest}
 
The \ac{cpd} identifies each rank-1 term only up to a per-component scalar.
The per-iteration normalization in~\eqref{eq:norm} fixes the norm of the allocation factors, so the ambiguity is absorbed into the signal factor $\hat{\bm{x}}_k$.
The effective channel gain is estimated by exploiting the discrete \ac{bpsk} structure of $\bm{s}_k$: as $[\bm{s}_k]_t\in\{\pm1\}$, squaring removes the modulation, $[\hat{\bm{x}}_k]_t^{2}\approx \tilde{h}_k^{2}$, so that averaging over all $T$ resources gives
\begin{equation}\label{eq:chest}
\hat{h}_k=\operatorname{sign}\!\big(\Re\{g_k^{*}[\hat{\bm{x}}_k]_{\mathcal{P}}\}\big)\,g_k,
\quad
g_k=\sqrt{\tfrac{1}{T}\textstyle\sum_{t=1}^{T}[\hat{\bm{x}}_k]_t^{2}},
\end{equation}
where $[\hat{\bm{x}}_k]_{\mathcal{P}}$ is the pilot resource used to resolve the sign of the square root. Unlike pilot-only estimation, \eqref{eq:chest} averages over the entire signal factor, thereby improving robustness to the non-coherent operation targeted here. The symbols are then recovered as 
\begin{equation}\label{eq:demod}
\hat{\bm{s}}_k=\hat{h}_k^{*}\,\hat{\bm{x}}_k,
\end{equation}
 followed by \ac{bpsk} demodulation $\hat{\bm{m}}_k=\operatorname{sign}(\Re\{\hat{\bm{s}}_k\})$.
 
\subsection{SIC Loop}
\label{sec:sic}
 
The detected bits are re-encoded through the transmit mapping, $\gamma(\hat{\bm{m}}_k)=(\tilde{\bm{c}}_{k,1},\dots,\tilde{\bm{c}}_{k,L})$, and a component is declared \emph{valid} when the re-encoded codewords match those returned by the projected decomposition across all modes, i.e., $\tilde{\bm{c}}_{k,\ell}=\hat{\bm{c}}_{k,\ell},\ \forall\ell$. As a residual (noise-only) component re-encodes to an inconsistent tuple with high probability, this test acts as an implicit error-detection mechanism, taking over the role
that an \ac{fec} code would otherwise play. \footnote{The codebook sizes govern the reliability of this check: with small codebooks, i.e., few distinct access patterns, an invalid component may satisfy the consistency test by chance, raising the false-alarm probability. Since a random tuple matches across all modes with probability $1/\prod_{\ell}|\mathcal{C}_\ell|$, larger codebooks make such coincidental validations increasingly unlikely.}
 
\begin{figure}[t]
    \centering
    \includegraphics[width=0.99\linewidth]{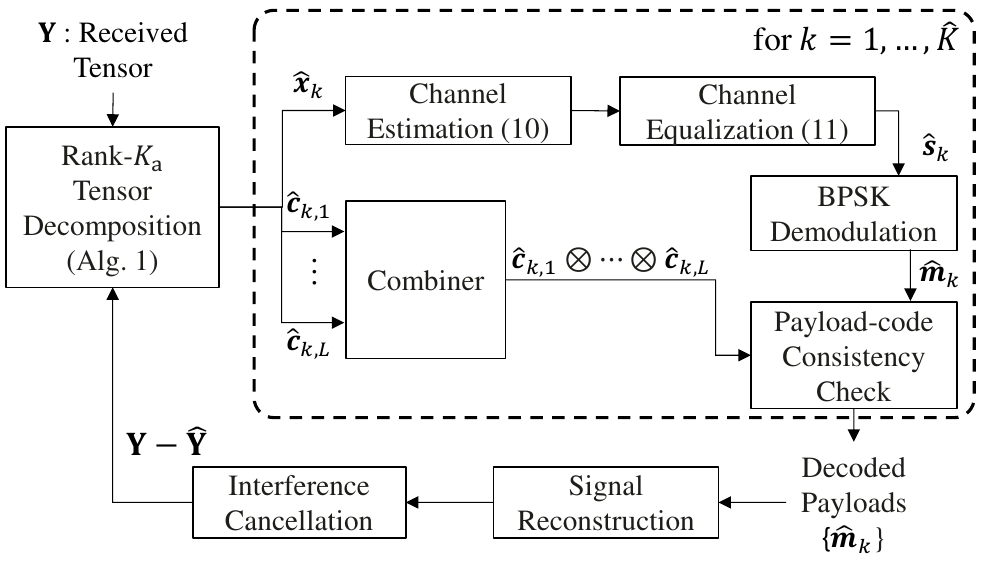}
    \caption{Iterative decoding process at the receiver.}
    \label{fig:receiver}
\end{figure}

Let $\mathcal{V}^{(i)}$ denote the set of components validated at \ac{sic} iteration $i$, initialized with $\mathbf{Y}^{(0)}=\mathbf{Y}$. To prevent noise propagation, the symbols of the validated components are re-modulated from their detected bits, and their tensor contributions are cancelled from the residual, 
\begin{equation}\label{eq:sic-recon}
\mathbf{Y}^{(i+1)}=\mathbf{Y}^{(i)}-\!\!\sum_{k\in\mathcal{V}^{(i)}}\!\!
\hat{h}_k\,\hat{\bm{s}}_k\otimes\hat{\bm{c}}_{k,1}\otimes\cdots\otimes\hat{\bm{c}}_{k,L}.
\end{equation}
The constrained \ac{cpd} of Algorithm~\ref{alg:als} is then re-applied to the residual $\mathbf{Y}^{(i+1)}$ with the rank reduced to $\hat{K}-\sum_{j\le i}|\mathcal{V}^{(j)}|$, exposing weaker or previously masked components. The loop stops if no new component is validated, or after at most $I_{\rm SIC}$ iterations. The receiver scheme is summarized in Fig.~\ref{fig:receiver}.

\section{Numerical Results}
\label{sec:numerical_results}

\Ac{tbra} is compared against the baseline \ac{saloha} and \ac{crdsa}, as well as their \ac{sic}-enhanced versions. 

\subsection{Simulation Setup}
We evaluate the proposed scheme over a random access frame with $N=128$ access occasions. Each active user transmits a $16$-bit \ac{aiot} identifier and one pilot symbol using \ac{bpsk}, resulting in $T=17$ resource elements per occasion. The channel follows the quasi-static Rayleigh model, with an average received \ac{snr} of $5$~dB. 
For all the \ac{sic}-enhanced schemes, we set $I_{\rm SIC}=100$.

\emph{Occasion Selection.}
In the following, \ac{tbra} configuration $(N_1,\ldots,N_L)$-$(C_1,\ldots,C_L)$ indicates a tensor of mode sizes $N_1,\ldots,N_L$ and weights, i.e., the number of ones per codeword, $(C_1,\ldots,C_L)$. 
The $16$-bit payload is mapped to the tuple $(\bm{c}_{k,1},\dots,\bm{c}_{k,L})$ by partitioning the bit sequence into segments and associating each segment with a codeword index. Specifically, letting $\bm{m}_{\ell,k}$ denote the $\ell$-th bit segment, the corresponding index is obtained as 
\begin{equation}\label{eq:codemap} 
i_{\ell,k} = \mathrm{bin2int}(\bm{m}_{\ell,k}) \bmod |\mathcal{C}_\ell|, 
\end{equation}
and the transmitted vector $\bm{c}_{k,\ell}$ is selected as the $i_{\ell,k}$-th codeword in $\mathcal{C}_\ell$.
This mapping ensures a structured and reproducible assignment of payloads to access patterns, while maintaining a sufficiently large space of codeword combinations. The same map~\eqref{eq:codemap} and reconstruction mechanism are applied to the compared state-of-the-art schemes, with the number of segments corresponding to the number of repetitions.

\emph{ALS Initialization.} The factors are initialized to reflect the signal structure of each mode. 
For the signal mode ($\ell =0$), each column is drawn as a random \ac{bpsk} sequence transmitted over a Rayleigh fading channel.
as
\begin{equation}\label{eq:init-x}
[\hat{\bm{x}}_k]_t =
\begin{cases}
\tilde{h}_k, & t = \mathcal{P},\\
\tilde{h}_k\,\varepsilon_{k,t}, & \text{otherwise},
\end{cases}
\end{equation}
with $\tilde{h}_k \sim \mathcal{CN}(0,1)$, and  $\varepsilon_{k,t}\in\{\pm1\}$, so that the pilot resources carry the gain $\tilde{h}_k$ and the data resources carry a uniformly random antipodal symbol, with $\varepsilon_{k,t}$ independent and equiprobable. 
Each codebook-mode factor is instead initialized as a non-negative uniform random vector normalized to the codeword norm, which places the initial estimates on the sphere of radius $\|\bm{c}\|_2$, $\bm{c}\in\mathcal{C}_\ell$, matching the constant Hamming weight of each codeword in $\mathcal{C}_\ell$.
To speed up convergence and improve the algorithm stability, \ac{als} parameters are adapted to the normalized load
\begin{equation}\label{eq:load}
\eta = \min\!\Big(\frac{\hat{K}}{R_{\rm id}},\,0.95\Big),
\end{equation}
which approaches $0$ when few users are active, and its cap as the rank nears the identifiability limit $R_{\rm id} = \min(T,N_1,\dots,N_L)$.
We set the number of iterations
\begin{equation}\label{eq:sched}
I_{\rm ALS}=\Big\lceil\frac{I_{\rm ALS}^{\rm min}}{1-\eta}\Big\rceil,
\quad
I_{\rm W}=\lceil 0.1\,(1+\eta)\,I_{\rm ALS}\rceil,
\end{equation}
respectively, and $\lambda=(1+\eta)\sigma_w^2$.
Discrete projection onto $\mathcal{C}_\ell$ is performed every $I_{\rm D}=40$ \ac{als} iterations, with $I_{\rm F}=5$. The minimum number of \ac{als} iterations is set to $I_{\rm ALS}^{\rm min}=50$.
Hence, low-rank decompositions, such as the late \ac{sic} passes, after most users have been cancelled, converge in few sweeps, whereas near the identifiability limit, the budget, the warm-up, and the regularization all increase to counter the slower convergence of the Khatri-Rao Gram matrix.

\subsection{Performance Results}

Upon receiving the uplink transmissions, the receiver produces a list $\hat{\mathcal{L}}=\{\hat{\bm{m}}_k:\,k=1,\ldots,\hat{K}_{\rm a}\}$ of $\hat{K}_{\rm a}$ decoded messages, which ideally coincides with the list $\mathcal{L}$ of transmitted ones. We measure accuracy through the \ac{mer}, defined as the expected fraction of transmitted messages that are not correctly recovered,
\begin{equation}\label{eq:mer}
\mathrm{MER} = \mathbb{E}\!\left[\frac{|\mathcal{L}\setminus\hat{\mathcal{L}}|}{K_{\rm a}}\right].
\end{equation}

\begin{figure}[t]
    \centering
    \setlength\fwidth{0.85\columnwidth}
    \setlength\fheight{0.6\columnwidth}
    \definecolor{TBRA844222color}{RGB}{214, 39, 40}    
\definecolor{TBRA16842color}{RGB}{140, 86, 75}    
\definecolor{TBRA844422color}{RGB}{148, 103, 189}  

\begin{tikzpicture}

\begin{axis}[%
    width=\fwidth,
    height=\fheight,
    at={(0\fwidth,0\fheight)},
    scale only axis,
    ylabel style={font=\scriptsize},
    xlabel style={font=\scriptsize},
    xmin=16,
    xmax=320,
    xlabel style={font=\color{white!15!black}},
    xlabel={Number of Users $K_{\rm a}$},
    ymin=0,
    ymax=1,
    yminorticks=true,
    ylabel style={font=\color{white!15!black}},
    ylabel={MER},
    axis background/.style={fill=white},
    xmajorgrids,
    ymajorgrids,
    xtick={32, 64, ..., 320, 320},
    ytick={0.2, 0.4, ..., 1},
    tick label style={font=\scriptsize},
    legend style={at={(0.365,0.01)}, 
                fill opacity=0.6,
                draw opacity=1,   
                text opacity=1,
                anchor=south west,
                legend cell align=left,
                align=left,
                font=\scriptsize,
                draw=white!15!black}
]

\addplot [color=TBRA16842color, dashed, very thick, mark=x, mark size=2, mark options={solid}]
table {%
16 0.304230769230769
32 0.336538461538462
64 0.544471153846154
96 0.839342948717949
128 0.927584134615385
160 0.956730769230769
192 0.967548076923077
224 0.974072802197802
256 0.981219951923077
288 0.991219951923077
320 1
};
\addlegendentry{TBRA $(16,8)-(4,2)$}

\addplot [color=TBRA16842color, dotted, very thick, mark=x, mark size=2, mark options={solid}]
table {%
16 0.0600961538461538
32 0.0733173076923077
64 0.207932692307692
96 0.487451923076923
128 0.645432692307692
160 0.767548076923077
192 0.839342948717949
224 0.885645604395604
256 0.906099759615385
288 0.933888888888889
320 0.946125
};
\addlegendentry{TBRA $(16,8)-(4,2)$ genie}

\addplot [color=TBRA16842color, solid, very thick, mark=x, mark size=2, mark options={solid}]
table {%
16 0.194807692307692
32 0.209134615384615
64 0.275841346153846
96 0.391025641025641
128 0.545673076923077
160 0.767548076923077
192 0.851762820512821
224 0.891311813186813
256 0.926832932692308
288 0.953888888888889
320 1
};
\addlegendentry{TBRA $(16,8)-(4,2)$ w/ SIC}

\addplot [color=TBRA844222color, dashed, very thick, mark=star, mark size=2, mark options={solid}]
table {%
16 0.277083333333333
32 0.30625
64 0.386979166666667
96 0.543402777777778
128 0.712239583333333
160 0.819166666666667
192 0.874652777777778
224 0.905803571428571
256 0.928776041666667
288 0.942584325396825
320 0.95546875
};
\addlegendentry{TBRA $(8,4,4)--(4,2,2)$}

\addplot [color=TBRA844222color, dotted, very thick, mark=star, mark size=2, mark options={solid}]
table {%
16 0.0416666666666667
32 0.0822916666666667
64 0.149479166666667
96 0.260763888888889
128 0.469791666666667
160 0.603125
192 0.666666666666667
224 0.74375
256 0.791015625
288 0.834821428571428
320 0.871875
};
\addlegendentry{TBRA $(8,4,4)-(2,2,2)$ Genie}

\addplot [color=TBRA844222color, solid, very thick, mark=star, mark size=2, mark options={solid}]
table {%
16 0.185416666666667
32 0.198958333333333
64 0.209895833333333
96 0.263541666666667
128 0.3040625
160 0.369375
192 0.435069444444444
224 0.514136904761905
256 0.592838541666667
288 0.674479166666667
320 0.72984375
};
\addlegendentry{TBRA $(8,4,4)-(2,2,2)$ w/ SIC}

\addplot [color=TBRA844422color, dashed, very thick, mark=triangle, mark size=2, mark options={solid}]
table {%
16 0.227083333333333
32 0.288541666666667
64 0.396875
96 0.614583333333333
128 0.779427083333333
160 0.866458333333333
192 0.9125
224 0.933928571428571
256 0.953645833333333
288 0.962425595238095
320 0.970424107142857
};
\addlegendentry{TBRA $(8,4,4)-(4,2,2)$}

\addplot [color=TBRA844422color, dotted, very thick, mark=triangle, mark size=2, mark options={solid}]
table {%
16 0.03125
32 0.0489583333333333
64 0.1046875
96 0.249652777777778
128 0.476197916666667
160 0.617708333333333
192 0.679375
224 0.735119047619048
256 0.792578125
288 0.834077380952381
320 0.865178571428571
};
\addlegendentry{TBRA $(8,4,4)-(4,2,2)$ genie}

\addplot [color=TBRA844422color, solid, very thick, mark=triangle, mark size=2, mark options={solid}]
table {%
16 0.122916666666667
32 0.148958333333333
64 0.1828125
96 0.25
128 0.302864583333333
160 0.381666666666667
192 0.468923611111111
224 0.582589285714286
256 0.69453125
288 0.76500496031746
320 0.825111607142857
};
\addlegendentry{TBRA $(8,4,4)-(4,2,2)$ w/ SIC}

\legend{}
\end{axis}

\begin{axis}[%
    width=\fwidth,
    height=\fheight,
    at={(0\fwidth,0\fheight)},
    scale only axis,
    xmin=1,
    xmax=100,
    xtick={},
    ytick={},
    xticklabels={{}, {}, {},{}},
    yticklabels={},
    xtick style = {draw=none},
    ytick style = {draw=none},
    ymin=0.0001,
    ymax= 1,
    legend style={
        /tikz/every even column/.append style={column sep=0.2cm},
        at={(0.59, 0.02)}, 
        anchor=south, 
        draw=white!80!black, 
        font=\scriptsize,
        fill opacity=0.8
        },
    legend columns=2,
]

\addplot [only marks, very thick, mark size=2, mark=x, mark options={solid, TBRA16842color}] table[row sep=crcr] {%
1	-5 \\
};
\addlegendentry{TBRA $(16,8)-(4,2)$}

\addplot [color=black, dashed, very thick]
  table[row sep=crcr]{%
1	-5\\
};
\addlegendentry{w/o SIC}

\addplot [only marks, very thick, mark size=2, mark=star, mark options={solid, TBRA844222color}] table[row sep=crcr] {%
1	-5 \\
};
\addlegendentry{TBRA $(8,4,4)-(2,2,2)$}

\addplot [color=black, dotted, very thick]
  table[row sep=crcr]{%
1	-5\\
};
\addlegendentry{Genie (w/o SIC) }

\addplot [only marks, very thick, mark size=2, mark=triangle, mark options={solid, TBRA844422color}] table[row sep=crcr] {%
1	-5 \\
};
\addlegendentry{TBRA $(8,4,4)-(4,2,2)$}

\addplot [color=black, solid, very thick]
  table[row sep=crcr]{%
1	-5\\
};
\addlegendentry{w/ SIC}

\end{axis}

\end{tikzpicture}
    \caption{MER versus the number of transmitting users $K_{\rm a}$.}
    \label{fig:pmissTBRA}
\end{figure}
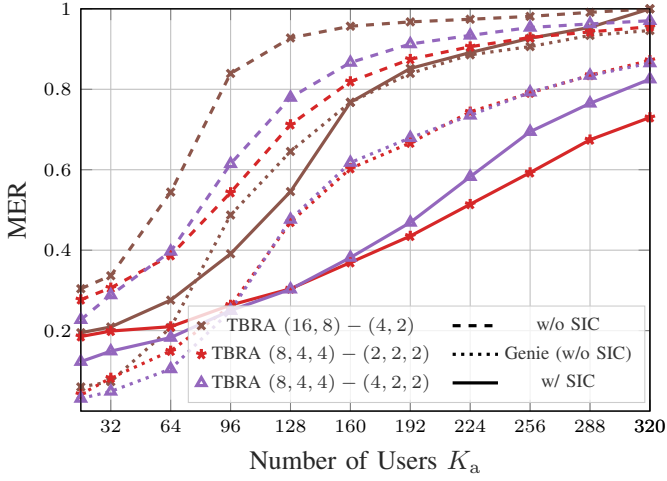

Fig.~\ref{fig:pmissTBRA} reports the \ac{mer} as a function of the number of active users $K_{\rm a}$ for different \ac{tbra} configurations, each shown for the blind receiver, the \ac{sic} receiver, and a genie-aided receiver in which the occasions selected by every user, i.e., the codewords of the modes $\ell>0$, are known. A first observation is the large gap between the two-mode configuration $(16,8)$-$(4,2)$ and the three-mode configurations $(8,4,4)$-$(2,2,2)$ and
$(8,4,4)$-$(4,2,2)$. Consistently with tensor identifiability theory, distributing the allocation across more modes improves the conditioning of the decomposition and thus the component recovery with \ac{als}~\cite[Th. 1]{Decurninge2021Tensor},~\cite{Chiantini2014An}. 
The genie-aided receiver reported here is without \ac{sic}. Consequently, although accurate at low load, it is surpassed by the \ac{sic} receiver as $K_{\rm a}$ grows, as by iteratively removing the validated users and re-decomposing the residual at reduced rank, \ac{als} is performed on lower rank tensors. Comparing the two weights configurations, the genie curves of $(2,2,2)$ and $(4,2,2)$ nearly coincide, so the number of selected occasions does not affect identifiability; under \ac{sic}, however, the lighter $(2,2,2)$ weight attains a lower \ac{mer} at high load, since fewer ones per mode reduce the access occasion occupancy where the receiver is most stressed. We therefore adopt $(8,4,4)$-$(2,2,2)$ in the following.

For comparison with state-of-the-art \ac{mac}-layer schemes, we further report the normalized throughput, i.e., the fraction of messages successfully resolved per access occasion,
\begin{equation}\label{eq:throughput}
\bar{T} = G\,(1 - \mathrm{MER}),
\end{equation}
where $G = K_{\rm a}/N$ is the normalized offered traffic.

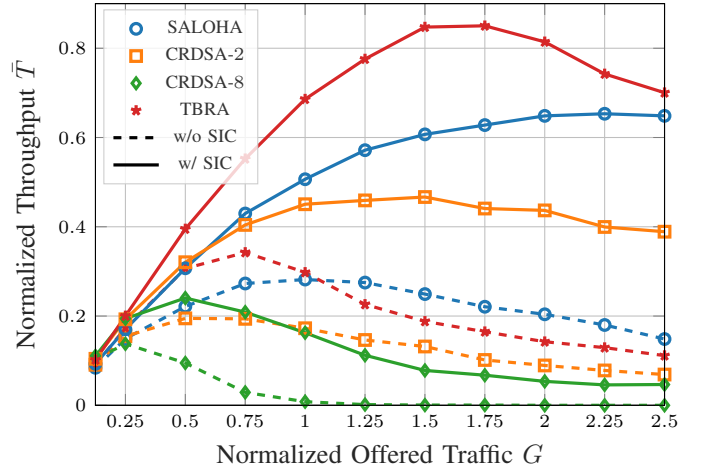
\begin{figure}[t]
    \centering
    \setlength\fwidth{0.85\columnwidth}
    \setlength\fheight{0.6\columnwidth}
    \definecolor{TBRA844222color}{RGB}{214, 39, 40}    
\definecolor{CRDSA8color}{RGB}{44, 160, 44}       
\definecolor{CRDSA2color}{RGB}{255, 127, 14}  
\definecolor{SALOHAcolor}{RGB}{31, 119, 180}     
\begin{tikzpicture}

\begin{axis}[%
    width=\fwidth,
    height=\fheight,
    at={(0\fwidth,0\fheight)},
    scale only axis,
    ylabel style={font=\scriptsize},
    xlabel style={font=\scriptsize},
    xmin=0.125,
    xmax=2.5,
    xlabel style={font=\color{white!15!black}},
    xlabel={Normalized Offered Traffic $G$},
    ymin=0.0,
    ymax=0.9,
    ylabel style={font=\color{white!15!black}},
    ylabel={Normalized Throughput $\bar{T}$},
    axis background/.style={fill=white},
    xmajorgrids,
    ymajorgrids,
    xtick={0.25, 0.5, ..., 2.5},
    tick label style={font=\scriptsize},
    legend style={at={(0.01,0.84)}, 
                fill opacity=0.6,
                draw opacity=1,   
                text opacity=1,
                anchor=south west,
                legend cell align=left,
                align=left,
                font=\scriptsize,
                draw=white!15!black},
    legend columns=2,
]

\addplot [color=SALOHAcolor, dashed, very thick, mark=o, mark size=2, mark options={solid}]
table{
0.125 0.0841145833333333
0.25 0.14921875
0.5 0.221354166666667
0.75 0.272916666666667
1 0.281770833333333
1.25 0.275
1.5 0.24921875
1.75 0.220833333333334
2 0.20390625
2.25 0.179966517857143
2.5 0.148482142857143
};
\addlegendentry{SALOHA}

\addplot [color=CRDSA2color, dashed, very thick, mark=square, mark size=2, mark options={solid}]
table {%
0.125 0.0908854166666667
0.25 0.1546875
0.5 0.194791666666667
0.75 0.19375
1 0.17265625
1.25 0.146354166666667
1.5 0.131770833333333
1.75 0.1015625
2 0.0893229166666667
2.25 0.078125
2.5 0.0689174107142856
};
\addlegendentry{CRDSA-$2$}

\addplot [color=CRDSA8color, dashed, very thick, mark=diamond, mark size=2, mark options={solid}]
table {%
0.125 0.103125
0.25 0.137760416666667
0.5 0.09484375
0.75 0.0290104166666667
1 0.00807291666666665
1.25 0.00156249999999997
1.5 0.000520833333333248
1.75 0.000520833333333331
2 0
2.25 0
2.5 0
};
\addlegendentry{CRDSA-$8$}

\addplot [color=TBRA844222color, dashed, very thick, mark=star, mark size=2, mark options={solid}]
table{
0.125 0.0903645833333333
0.25 0.1734375
0.5 0.306510416666667
0.75 0.342447916666667
1 0.297760416666667
1.25 0.226041666666667
1.5 0.188020833333333
1.75 0.16484375
2 0.142447916666667
2.25 0.129185267857143
2.5 0.111328125
};
\addlegendentry{TBRA $(8,4,4)-(2,2,2)$}

\addplot [color=SALOHAcolor, solid, very thick, mark=o, mark size=2, mark options={solid}]
table{
0.125 0.09296875
0.25 0.1703125
0.5 0.30703125
0.75 0.4296875
1 0.506510416666667
1.25 0.571614583333333
1.5 0.60703125
1.75 0.627864583333333
2 0.648489583333333
2.25 0.653203125
2.5 0.64854910714286
};
\addlegendentry{SALOHA w/ SIC}

\addplot [color=CRDSA2color, solid, very thick, mark=square, mark size=2, mark options={solid}]
table {%
0.125 0.104166666666667
0.25 0.192708333333333
0.5 0.320572916666667
0.75 0.404166666666667
1 0.450520833333333
1.25 0.459114583333333
1.5 0.466666666666667
1.75 0.440885416666667
2 0.43671875
2.25 0.399553571428572
2.5 0.388950892857143
};
\addlegendentry{CRDSA-$2$ w/ SIC}

\addplot [color=CRDSA8color, solid, very thick, mark=diamond, mark size=2, mark options={solid}]
table {%
0.125 0.110677083333333
0.25 0.194270833333333
0.5 0.240364583333333
0.75 0.208802083333333
1 0.162604166666667
1.25 0.112083333333333
1.5 0.0781250000000001
1.75 0.0674479166666665
2 0.0536458333333334
2.25 0.0457589285714284
2.5 0.0465959821428572
};
\addlegendentry{CRDSA-$8$ w/ SIC}

\addplot [color=TBRA844222color, solid, very thick, mark=star, mark size=2, mark options={solid}]
table{
0.125 0.101822916666667
0.25 0.200260416666667
0.5 0.395052083333333
0.75 0.55234375
1 0.6859375
1.25 0.77578125
1.5 0.847395833333333
1.75 0.850260416666667
2 0.814322916666667
2.25 0.742421875
2.5 0.700390625
};
\addlegendentry{TBRA $(8,4,4)-(2,2,2)$ w/ SIC}

\legend{}
\end{axis}

\begin{axis}[%
    width=\fwidth,
    height=\fheight,
    at={(0\fwidth,0\fheight)},
    scale only axis,
    xmin=1,
    xmax=100,
    xtick={},
    ytick={},
    xticklabels={{}, {}, {},{}},
    yticklabels={},
    xtick style = {draw=none},
    ytick style = {draw=none},
    ymin=0.0001,
    ymax= 1,
 legend style={
        /tikz/every even column/.append style={column sep=0.2cm},
        at={(0.15, 0.55)}, 
        anchor=south, 
        draw=white!80!black, 
        font=\scriptsize,
        fill opacity=0.8
        }
]

\addplot [only marks, very thick, mark size=2, mark=o, mark options={solid, SALOHAcolor}] table[row sep=crcr] {%
1	-5 \\
};
\addlegendentry{SALOHA}

\addplot [only marks, very thick, mark size=2, mark=square, mark options={solid, CRDSA2color}] table[row sep=crcr] {%
1	-5 \\
};
\addlegendentry{CRDSA-$2$}

\addplot [only marks, very thick, mark size=2, mark=diamond, mark options={solid, CRDSA8color}] table[row sep=crcr] {%
1	-5 \\
};
\addlegendentry{CRDSA-$8$}

\addplot [only marks, very thick, mark size=2, mark=star, mark options={solid, TBRA844222color}] table[row sep=crcr] {%
1	-5 \\
};
\addlegendentry{TBRA}

\addplot [color=black, dashed, very thick]
  table[row sep=crcr]{%
1	-5\\
};
\addlegendentry{w/o SIC}

\addplot [color=black, solid, very thick]
  table[row sep=crcr]{%
1	-5\\
};
\addlegendentry{w/ SIC}

\end{axis}

\end{tikzpicture}
    \caption{Normalized throughput versus the normalized offered traffic $G$.}
    \label{fig:throughput}
\end{figure}

Fig.~\ref{fig:throughput} shows the normalized throughput versus $G$. All baselines are simulated with capture and, where indicated, \ac{sic}, so that intra-access occasion collisions may already be partially resolved. Without \ac{sic} (dashed), \ac{saloha} peaks at $\bar{T}\approx0.28$ near $G=1$, whereas increasing the \ac{crdsa} degree raises the access occasion occupancy and is detrimental. Notably, \ac{crdsa} surpasses \ac{saloha} only at very low load: with collisions already resolved within the access occasion by capture, the replica diversity is largely redundant while its occupancy cost remains.
This contrast is central to the proposed design: plain replication multiplies the load without adding separability, whereas \ac{tbra} structures the repetitions so that the receiver can exploit them to separate users.
With \ac{sic} (solid), \ac{tbra} dominates regardless of the traffic load, peaking at $\bar{T}\approx0.85$ at $G\approx1.75$, showing about a $30\%$ improvement on the best baseline.

\subsection{Complexity Considerations \& AIoT Constraints}
\label{sec:complexity}
At the \textit{transmitter} side, each device maps its $B$-bit payload to a codeword tuple through $\gamma(\cdot)$ and transmits the same \ac{bpsk} vector over its $|\mathcal{S}_k|$ occasions, for $\mathcal{O}(B+dT)$, with no channel coding and no training. This matches the transmitter of \ac{crdsa}, as the tensor structure is induced solely by the choice of occasions in $\gamma(\cdot)$ and adds no complexity.
\ac{tbra} thus retains the minimal transmitter required by \ac{aiot} devices.

On the \textit{reader} side, instead, the \ac{tbra} receiver reuses the \ac{sic} control flow of \ac{crdsa}, adding only the constrained \ac{cpd} that separates users superimposed within an access occasion, i.e., the collisions that repetition schemes cannot resolve. The extra cost, $\mathcal{O}\big(I_{\rm ALS}(\hat{K}^2 TN + L\hat{K}^3)\big)$ per \ac{sic} pass, is polynomial and carried entirely at the reader, which typically has less stringent constraints in energy or hardware complexity, being therefore affordable in \ac{aiot} settings. The estimation and equalization procedures are shared with the \ac{saloha} and \ac{crdsa} baselines and add no scheme-specific cost.

\section{Conclusions} 
\label{sec:conclusions}
This letter introduced \ac{tbra}, a tensor-based grant-free random access scheme for \ac{aiot} Msg1 transmissions under stringent hardware constraints. By mapping each payload to a structured multi-dimensional access pattern, the design induces a separable tensor model that a constrained decomposition with \ac{sic} resolves, jointly performing activity detection, channel estimation, and decoding with a single-antenna receiver, low-order modulation, and no \ac{fec}. Numerical results show substantial reliability gains over slotted ALOHA-based schemes while preserving an ultra-low-complexity transmitter.

\balance

\bibliographystyle{IEEEtran}
\bibliography{IEEEabrv,references}

\end{document}